\documentclass[a4paper,fleqn,usenatbib]{mnras}
\usepackage[T1]{fontenc}
\usepackage{ae,aecompl}

\usepackage{amsmath}
\usepackage{graphicx}
\usepackage{lscape}
\usepackage{indentfirst}
\usepackage{xspace}

\usepackage{amssymb}
\usepackage{color}
\usepackage[flushleft]{threeparttable}

\newcommand {\bc}{\begin {center}}
\newcommand {\ec}{\end {center}}
\newcommand {\be}{\begin {equation}}
\newcommand {\ee}{\end {equation}}
\newcommand {\bea}{\begin {eqnarray}}
\newcommand {\eea}{\end {eqnarray}}

\newcommand {\comment}[1]{}

\def\lbar {\lambda\hskip-5pt\raise3pt\hbox {--}}
\def\lbr {\lambda\raise2pt\hbox {\hskip-4pt{\scriptsize --}}_\C}

\def\vv {{\rm v}}

\title[Mean opacities of a strongly magnetized high temperature plasma]
{Mean opacities of a strongly magnetized high temperature plasma}
\author[V.F.~Suleimanov et al.] 
{Valery~F.~Suleimanov,$^{1}$\thanks{E-mail: suleimanov@astro.uni-tuebingen.de (VFS)}  
 Alexander~A.~Mushtukov,$^{2,3}$
  Igor Ognev,$^4$ \newauthor
 Victor A. Doroshenko,$^{1}$
  Klaus Werner$^1$
 \\ 
$^{1}$Institut f\"{u}r Astronomie und Astrophysik, Kepler Center for Astro and
Particle Physics, Universit\"{a}t T\"{u}bingen, \\ Sand 1, 72076 T\"{u}bingen,
Germany\\
$^2$Astrophysics, Department of Physics, University of Oxford, Denys Wilkinson Building, Keble Road, Oxford OX1 3RH, UK \\
$^3$Leiden Observatory, Leiden University, NL-2300RA Leiden, The Netherlands\\
$^4$P. G. Demidov Yaroslavl State University, Sovietskaya 14, 150003 Yaroslavl, Russia
} 

\pubyear{2022}

\begin{document}
\label{firstpage}
\pagerange{\pageref{firstpage}--\pageref{lastpage}}
\maketitle

\begin{abstract}
Geometry and dynamical structure of emission regions in accreting pulsars are shaped by the interplay between gravity, radiation, and strong magnetic field, which significantly affects the opacities of a plasma and radiative pressure under such extreme conditions. 
Quantitative  consideration of magnetic plasma opacities is, therefore, an essential ingredient 
 of any self-consistent modeling of 
emission region structure of X-ray pulsars. 
 We present results of computations of the Rosseland and Planck mean opacities of a strongly magnetized plasma with a simple chemical composition,
namely the solar hydrogen/helium  mix. 
We consider all relevant specific 
opacities of the magnetized plasma including vacuum polarization  effect
and contribution of electron-positron pairs where the pair number density is computed in the thermodynamic equilibrium approximation.  The magnetic Planck mean opacity determines the radiative cooling of an optically thin strongly 
 magnetized plasma. 
 It is by factor of three 
  smaller than non-magnetic Planck opacity at 
 $k_{\rm B}T < 0.1\,E_{\rm cyc}$ and increases by a factor of  10$^2$ -- 10$^4$ 
 at $k_{\rm B}T > 0.3\,E_{\rm cyc}$ due to 
 cyclotron thermal processes.  
 We  propose  a simple approximate expression  which has sufficient accuracy for the magnetic  Planck opacity description.
 We provide the Rosseland opacity 
  in a tabular form computed in the temperature range 1 -- 300 keV, magnetic field range $3 \times 10^{10} - 10^{15}$\,G, and a broad range of plasma densities. 
  We demonstrate that  the scattering  on the electron-positron  pairs increases the Rosseland opacity drastically at temperatures 
$>50$\,keV in the case of mass densities typical for accretion channel in X-ray pulsars. 
 
\end{abstract}



\section{Introduction}
\label{sec:Intro}
Emission regions of X-ray pulsars 
 (XRPs, see e.g., \citealt{WSH83,2022arXiv220414185M}) and pulsing ultraluminous X-ray sources (pULXs) confine high temperature and high density plasma threaded by ultra-strong magnetic fields \citep{2014Natur.514..202B, 
Kaaretetal.17, Fabrika.etal:21}. 
Despite many attempts over the last few decades, there are no self-consistent physical models
accurately describing the dynamical structure of  the emission regions in these objects and their observed X-ray spectra. 
Two main directions of research can be identified here, i.e. attempts aiming to describe observed spectra based on some simplified assumptions regarding emission region geometry and dynamical structure, and modeling aimed to justify such assumptions from first principles.
As an example of first approach one can quote models by \cite{BW07} \cite{Farinelli.etal:16} which were successfully used for description of the observed spectra of highly-luminous XRPs \citep[see e.g.][]{Wolff.etal:16, CH21}, albeit using a rather extended set of free parameters  and at times ignoring features like the energy conservation law \citep{2021A&A...656A.105T}. 
A similar approach was used to model spectra of low-luminous objects \citep{Mushtukov.etal:21, SL.etal:21}, which were recently found to exhibit two-humped X-ray spectra \citep{2019MNRAS.483L.144T, 2019MNRAS.487L..30T}.

On the other hand, several attempts have been made to model  
 actual physical conditions in the accreting regions, especially in accretion columns. 
For instance, the pioneering work by \citet{BS76} and many other related investigations \citep[see e.g.][]{1981A&A....93..255W, LS88}, 
including two- and three-dimensional modeling of the accretion columns \citep{Postnov.etal:15, TO17, G21}. 
Note that understanding of 
 physical conditions in the emitting region of XRPs is essential 
to constrain the maximal possible luminosity of accretion columns, which is 
relevant for understanding the phenomenon of pULXs \citep{Mushtukovetal.15, 
Brice.etal:21}. 
In a first order approximation, the maximal luminosity depends on the optical depth across the column
\citep[see e.g.][]{LS88} and, therefore, on the plasma opacities which remain a major source of uncertainty in such modeling. Indeed, up to now only electron scattering with and without 
magnetic field  were taken into account. 
In particular, the reduction of the electron scattering cross-section in strong 
magnetic fields was used by \citet{Mushtukovetal.15} to explain the observed high luminosities of pULXs 
\citep[see also][]{Brice.etal:21}. 

Other 
 processes can, however, contribute to the high temperature opacity of plasma in a strong magnetic field. The simplest one is magnetic bremsstrahlung
\citep[see e.g. ][]{KPS83}. 
At high plasma temperatures 
creation  of electron-positron pairs due to 
photon-photon interactions 
comes into play \citep[see e.g. ][and references therein]{B99}. 
In presence of strong magnetic fields, also one-photon pair creation becomes relevant \citep[see e.g.][]{DH83}. 
Note that the number density of pairs can be significant in
an optically thick plasma in thermodynamic equilibrium especially 
 under the conditions of extremely strong magnetic fields
\citep[see e.g.][]{Mushtukovetal.19}. 
The increase of pair number density naturally leads to the increase of the opacity.
One of the aims of the current work is to compute the Rosseland mean opacity of a high temperature plasma in a strong magnetic field including all the mentioned processes and 
answer the question whether the aforementioned effects could be relevant for different types of accretion column models.   

 Here we provide simple means for calculation of the opacity of magnetized plasma and thus make a step towards the accurate calculations of thermal balance in radiative transfer models accounting for a strong external magnetic field, which was an issue for some of the previous investigations.
For instance, \citet{SL.etal:21} only considered non-magnetized hydrogen plasma emissivity due to bremsstrahlung to compute temperature structures of the overheated upper layers of strongly magnetized neutron star atmospheres heated
by accreting protons. We note, however, that plasma cooling and thermal balance could be different if additional cyclotron emission and  magnetic bremsstrahlung would be taken into account. The optically thin plasma cooling can be expressed in terms of the Planck mean
opacity. Its accurate computation for a high temperature magnetized plasma is the second aim of the paper. 

\section{Method}
\label{method}

To calculate the mean Planck $k(\rho,T,B)$ and Rosseland $\kappa(\rho,T,B)$ 
opacities of a strongly magnetized plasma at given plasma density $\rho$, temperature $T$,
and magnetic field strength $B$, we have to determine 
the plasma chemical composition, number densities of the plasma particles, 
and cross-sections of the elementary processes of  
photon interactions with the plasma particles, other photons, and magnetic field.
Here we assume that the plasma is the solar hydrogen/helium mix without heavy elements. We assume also that both
chemical elements are fully ionized at the considered temperatures $k_{\rm B}T > 1$\,keV, where $k_{\rm B}$ is
the Boltzmann constant. 

For computations of the number densities of the plasma particles we have to take into account 
that the Planck and Rosseland mean opacities are relevant at significantly different physical conditions. 
The Planck mean opacity describes the radiation loss rate of an optically thin plasma, 
$Q = k(\rho,T,B)B(T)$, where $B(T)=\sigma_{\rm SB}T^4/\pi$ is the integral Planck function, 
and $\sigma_{\rm SB}$ is the Stefan-Boltzmann constant.
Conversely,  the Rosseland mean opacity controls the radiation energy transport in optically thick plasma. 
Therefore, we assume  that the plasma is optically thick and in thermodynamic equilibrium when we compute
 the Rosseland opacity. In particular it means that the radiation intensity equals the Planck function 
at the considered temperature. We also assume  that the electron-positron pairs are in thermodynamic equilibrium 
and their number densities can be  computed here using this approximation. 
On the other hand, we cannot exclude physical conditions when the radiation field is still in equilibrium but the pair creation and annihilation is not. Furthermore, the Rosseland mean opacity can be useful for estimates of radiative acceleration in the upper layers of neutron stars \citep[see, e.g.][]{2015MNRAS.447.1847M}, i.e. under the assumption of a
significantly diluted radiation field. Therefore, we computed three sets of the Rosseland mean opacities
for several assumptions regarding the pairs.
First, we consider the case with pairs in thermodynamical equilibrium. For the second case pair number density is
assumed to be insignificant, but they can be created by photons assuming that the radiation field is described by the Planck function
at the considered temperature. Finally, we consider also the case where all processes involving the pairs are ignored. 

Pairs  certainly cannot be in thermodynamic equilibrium in an optically thin plasma, and we ignore pairs and all the processes 
connected with pair creations when we compute the Planck opacity. Details of number densities calculations are presented 
in Appendix\,{\ref{sec:numbers}}.

 We consider all relevant opacity sources in the high-temperature magnetized plasma to compute the Rosseland 
and Planck means. 
In particular, we consider specific opacities due to  electron scattering $\kappa_{\rm es}^j$ and
 free-free absorption $\kappa_{\rm ff}^j$ (continuum), as well as opacities in the cyclotron resonance and harmonics 
 $\kappa_{\rm cyc}^j$. Finally, we treat the processes of electron-positron pair creations by photons as an opacity. 
 We compute the continuum and cyclotron opacities separately in two normal modes, the extraordinary one,
 $X\,(j=1)$ and the ordinary one, $O\,(j=2)$. The opacities due to
the two-photon $\kappa_{\gamma\gamma}$ and one-photon pair creation  in a strong magnetic field $\kappa_{1\gamma}$ 
are computed without distinguishing the modes, and we just take them to be equal for both modes. 
 We emphasize that this recipe can also be considered an assumption of our model.

We use approximate expressions rather than precise treatment to simplify the calculations whenever possible. All specific opacities used in the calculations are described in Appendix\, \ref{sopac}.
The main approximations used are described here. First, we use the rarified plasma approximation, which means that the plasma frequency 
$\nu_{\rm p}=\sqrt{e^2 n_e/\pi m_e}$ is  significantly less than the photon frequency $\nu$, and we use the cold plasma 
approximation for the opacities below the electron cyclotron  frequency $\nu_{\rm B}=eB/2\pi m_{\rm e}c$.  The corresponding 
plasma  energy is $E_{\rm p} \approx 0.12\,\sqrt{n_{\rm e}/10^{25}}$\,keV, and the corresponding non-relativistic cyclotron energy is
 $E_{\rm cyc} = \hbar eB/m_{\rm e} c \approx 11.6\, B_{12}$\,keV.
Here and further on we use the definition $B_{12} = B/10^{12}$\,G, and, also, $n_{\rm e}$ instead of
the sum of the electron and positron number densities, $n_{\rm e} = n_{\rm e^-} +n_{\rm e^+}^{\rm B}$ 
when the Rosseland mean opacities are considered.

The specific total opacity $\kappa^j(E,\mu)$ 
at the photon energy $E$ depends on $\mu=\cos\theta$, where $\theta$ is the angle
 between the photon momentum and the magnetic field direction, and is computed as 
\be
      \kappa^j(\mu,E) = k^j(\mu,E)+\max(\kappa_{\rm es}^j,(1-\mathcal{P})\kappa_{\rm cyc}^j)+
      \kappa_{\gamma\gamma}+\kappa_{1\gamma}. 
\ee 
The first term is the true opacity
\be
      k^j(\mu,E) = \max(\kappa_{\rm ff}^j,\mathcal{P}\,\kappa_{\rm cyc}^j),
\ee 
which includes free-free absorption and thermal cyclotron emission.
Here we took into account the fact, that the continuum opacities and the cyclotron opacity are closely tied and 
have to be considered and computed  simultaneously. However, we describe both here as different processes 
in the framework of the adopted approximations. 
Therefore, we compare  $\kappa_{\rm es}^j$ with $(1-\mathcal{P})\kappa_{\rm cyc}^j$ and 
$\kappa_{\rm ff}^j$  with $\mathcal{P}\,\kappa_{\rm cyc}^j$
at every photon energy, and only take into account the largest opacity.

 The main processes of excitation and de-excitation of upper Landau levels are interactions with photons. 
On the other hand, de-excitation and excitation by thermal ions
also contribute to the total rate of excitation and de-excitation. 
Thus, the process where excitation is due to photon and de-excitation 
is due to collision with an ion must be considered as a true opacity.
The opposite process provides additional plasma cooling.
 The relative contribution of the thermal processes to the total cyclotron
opacity $\mathcal{P}$ can be expressed as  it was described by \citet{PMS80}, see also discussion 
in \citet{Potekhin14} 
\be \label{eps}
       \mathcal{P} = \frac{\nu_{\rm ei}}{\nu_{\rm ei}+\nu_{\rm er}} \approx \frac{\nu_{\rm ei}}{\nu_{\rm er}},
\ee
where $\nu_{\rm er}$ is the frequency of the electron-photon
collisions  and $\nu_{\rm ei}$ is the frequency of the electron-ion collisions. 
We used, however, the full $\mathcal{P}$ definition (the left side of Eq.\,\ref{eps}) in the computations.

The collision frequencies are determined as  \citep[see e.g.][]{vAL06}
\be
  \nu_{\rm er}\approx \frac{2e^2}{3\,\hbar^2\,m_{\rm e}\,c^3}\,E^2 \approx 1.446\times 10^{13}\,E^2_{\rm keV} 
\ee
\bea
     \nu_{ei}  \approx  \frac{4\pi}{3\sqrt{3}}\bar Z^2 e^4 \left(
\frac{2\pi}{m_e kT}\right)^{1/2} \frac{n_{\rm ion}}{E}
\left[1-\exp{\left(-\frac{E}{k_{\rm B}T}\right)}\right] \\
  \nonumber
  \approx 1.93 \times 10^{-10} \frac{n_{\rm ion}}{T_{\rm keV}^{1/2}E} \left[1-\exp{\left(-\frac{E}{k_{\rm B}T}\right)}\right]. 
\eea 
Therefore,
\be
       \mathcal{P}  \approx  1.3 \times 10^{-23} \frac{n_{\rm ion}}{T_{\rm keV}^{1/2}E^3} \left[1-\exp{\left(-\frac{E}
       {k_{\rm B}T}\right)}\right].
\ee
This value agrees to an order of magnitude with the estimation made by  \citet{AKL87}, i.e. given by their Eqs.\,(64) and (66). It is also similar to the value presented by \citet{Mushtukov.etal:21}, i.e. given by their Eq.\,(19), if we assume that $E=k_{\rm B}T=E_{\rm cyc}$. 
The estimates in both papers cited above were made, however, for the cyclotron line only, so our result is acceptable in a wider energy range.

It is also important to realize that mean opacities of a magnetized plasma depend on the angle between the magnetic field lines and 
the direction of the photon propagation.  We computed the Rosseland  mean for two such angles,
across and along magnetic field for both polarization normal modes. The Rosseland
mean across the field  is computed  as it was 
described by \citet{Mushtukovetal.15}
\be
     \kappa_{\perp}^j = \frac{\int\limits_0^\infty dB_E/dT\,dE}{\int\limits_0^\infty dB_E/dT\,dE
     \int\limits_0^\pi d\varphi \int\limits_0^1 3\mu_n^2[\pi \kappa_j(\mu, E)]^{-1}\,d\mu_n},     
\ee 
where $E$ is the photon energy, and $B_E$ is the Planck function. The angular integration occurs around the main axis
normal to the magnetic field direction, and $\varphi$ is the azimuthal angle, and $\mu_n$ is the cosine of the angle 
between the photon momentum and the main axis. We note that the specific total opacity $\kappa_j(E,\mu)$ 
at photon energy $E$ depends on the cosine $\mu=\cos\theta$, where $\theta$ is the angle
 between photon momentum and magnetic field direction. Both cosines are connected as 
 $\mu=\sqrt{1-\mu_n^2}\cos\varphi$. We note that $\kappa_j (\mu,E)$ depends also on plasma temperature 
 and density.
 
The Rosseland mean opacity along the magnetic field is computed more easily due to axial symmetry \citep{2015MNRAS.447.1847M}
 \be
     \kappa_{\parallel}^j = \frac{\int\limits_0^\infty dB_E/dT\,dE}{\int\limits_0^\infty dB_E/dT\,dE
     \int\limits_0^1 3\mu^2[\kappa_j(\mu, E)]^{-1}\,d\mu}.    
\ee 

The Planck mean opacities in the both directions are computed in a similar manner, but here we only considered the opacity 
along the field, because there is an obvious astrophysical 
application only for this case \citep[see e.g.][]{SL.etal:21}:
\be
     k_{\parallel}^j = \frac{\pi\int\limits_0^\infty B_E\,dE
     \int\limits_0^1 k_j(\mu, E)\,d\mu}{\sigma_{\rm SB}T^4},     
\ee 
with the same notation. We note, that the both mean opacities are computed in the
plasma rest-frame.

The mean opacities computed separately in two polarization modes  were summed using following rules
\be
         \frac{1}{\kappa}_{\perp,\parallel} = \frac{1}{2\kappa^{\rm X}_{\perp,\parallel}}+\frac{1}{2\kappa^{\rm O}_{\perp,\parallel}}\hspace{0.5cm}{\rm and} 
         \hspace{0.5cm} k_{\parallel} = \frac{k^{\rm X}_{\parallel}}{2}+\frac{k^{\rm O}_{\parallel}}{2}. 
 \ee
We mix the Rosseland opacities for both modes in equal proportions
because electron scattering effectively mixes the polarization modes in the optically thick plasma, where the Rosseland mean 
opacity has to be used. This is a consequence of the assumption of local thermodynamical equilibrium in the plasma, which is
optically thick in both modes. In particular, this assumption $I^j_E = B_E/2$ is used as the inner boundary condition for the radiation
transfer equations in the normal modes \citep{Shibanovetal.92}. Cooling of a plasma optically thin in both modes occurs
independently in both modes,  $Q^j = k^j_{\parallel}B(T)/2$, and the total cooling rate is reduced to 
$Q =( k^{\rm X}_{\parallel}/2+k^{\rm O}_{\parallel}/2)B(T)$. 

Integration over photon energies cannot be performed from zero to infinity in numerical computations. Therefore, we used
finite limits during the computations.  It is important to cover a wide band around the Planck function maximum at a given 
temperature. Therefore, we take the energy band for computing the mean opacities at given temperature between
 $0.01\,k_{\rm B}T$ and  $30\,k_{\rm B}T$. Values of the Planck function at the boundaries are at least by four orders of magnitude lower compared to the maximum, and this fact guarantees that the outside energy bands do not contribute significantly to the mean opacity.

\section{Results}

\begin{figure}
\centering
\includegraphics[angle=0,scale=0.99]{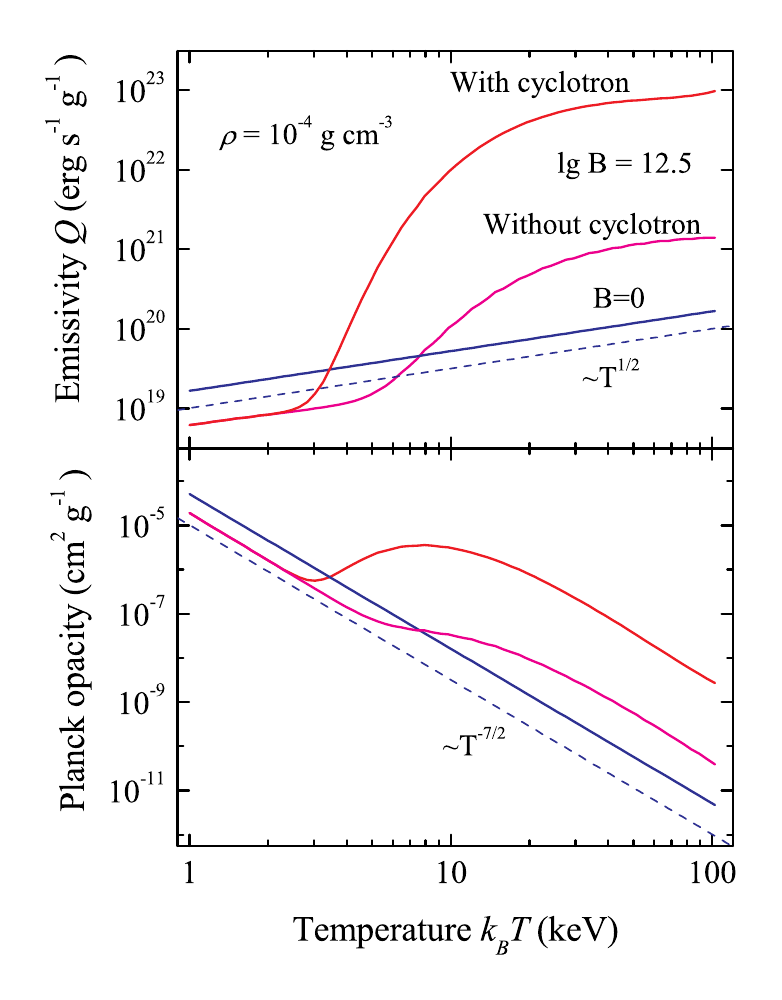}
\caption{\label{fig:p_comp}
Dependencies of the magnetized plasma emissivity (top panel) and the Planck mean opacity (bottom panel) 
on the plasma temperature 
for magnetic field strength $B = 10^{12.5}$\,G computed with and without contribution of the cyclotron emission.
The corresponding values computed for non-magnetized plasma are also shown. 
The plasma density is $\rho = 10^{-4}$\,g\,cm$^{-3}$.
}
\end{figure}

Thermal cyclotron emission significantly contributes to the Planck opacity and, therefore, to radiative energy losses 
at some temperatures, see Fig.\,\ref{fig:p_comp}. The Planck opacity of the magnetized plasma 
is lower by about a factor of two to three compared to 
the corresponding opacity of the non-magnetized plasma if the plasma temperature is much less than the cyclotron energy 
$(k_{\rm B}T \le 0.1\,E_{\rm cyc})$. The main reason is that 
the opacity in $X$-mode is depressed and only the $O$-mode contributes to the opacity under this condition.
The cyclotron thermal emission completely dominates at $kT > 0.3\,E_{\rm cyc}$ increasing the Planck opacity by approximately 
three orders of magnitude. We note, however, that the Planck opacity of the magnetized plasma increases even if we ignore
thermal cyclotron emission ($\mathcal P = 0$), which is due to the contribution of peaks in the Gaunt factor at the cyclotron harmonics \citep[see][]{PP76, SPavW10, Potekhin10}.

\begin{figure}
\centering
\includegraphics[angle=0,scale=0.99]{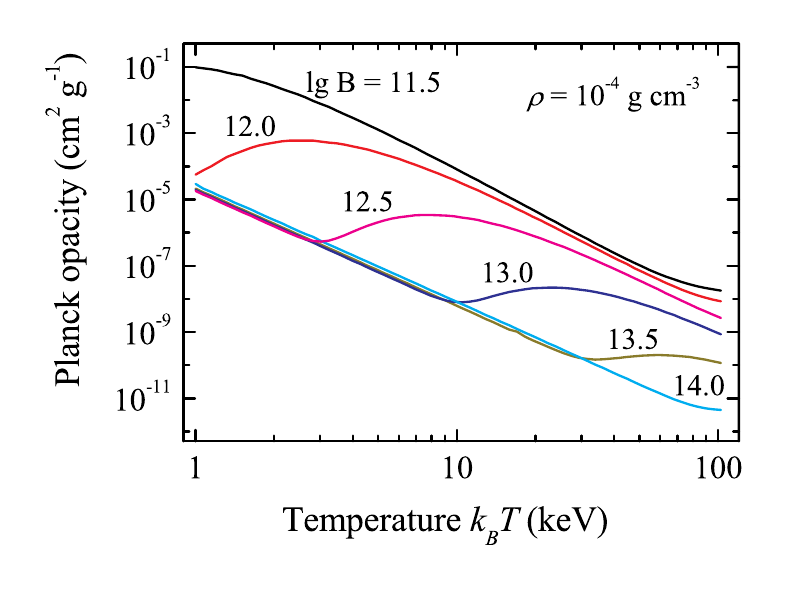}
\caption{\label{fig:p_varB}
Dependencies of the magnetized plasma  Planck mean opacities on the plasma temperature 
for various magnetic field strengths and 
plasma density $\rho = 10^{-4}$\,g\,cm$^{-3}$.
}
\end{figure}

\begin{figure}
\centering
\includegraphics[angle=0,scale=0.99]{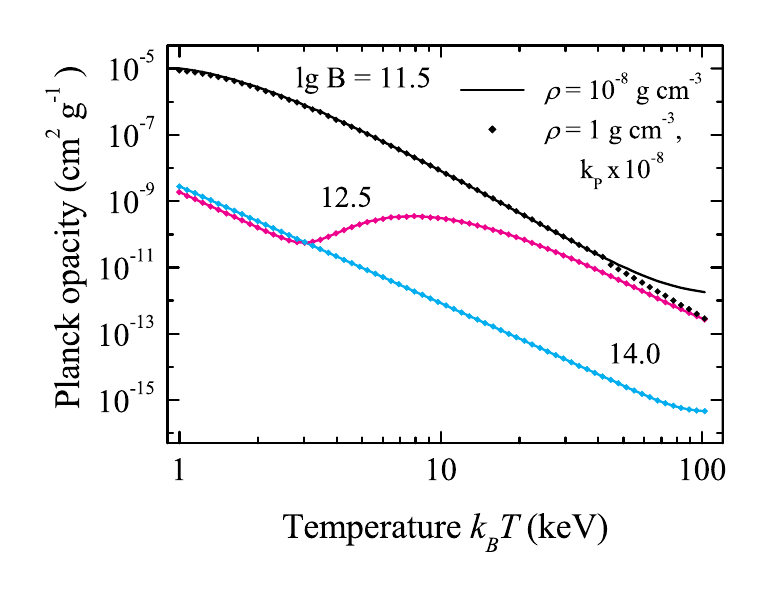}
\caption{\label{fig:p_rho}
Comparison of the Planck mean opacities computed for two different plasma densities,  $\rho = 10^{-8}$\,g\,cm$^{-3}$
and $\rho = 1$\,g\,cm$^{-3}$, and three various magnetic field strengths. The opacities computed for the higher density
were multiplied by a factor of $10^{-8}$. 
}
\end{figure}

The dependence of the Planck mean opacity on the magnetic field strength is shown in Fig.\,\ref{fig:p_varB}. It can be seen that the
Planck opacity properties are consistent with the description presented above. However, the contribution of thermal cyclotron emission
becomes relatively less important as the magnetic field strength increases. It is well known, that the Planck opacity 
of a non-magnetized plasma depends linearly on plasma density. The same is correct for a magnetized plasma,
see Fig.\,\ref{fig:p_rho}, where the normalized Planck opacities computed for $\rho = 10^{-8}$\,g\,cm$^{-3}$ and 
$\rho = 1$\,g\,cm$^{-3}$, and three various magnetic field strengths are shown. Note that there is some inconsistency at the lowest
magnetic field and the highest temperature, which is related to the fact that the approximations used for the cyclotron opacity at these plasma parameters start to fail here because the cyclotron harmonics strongly overlap at these conditions, and much more complicated 
computations are necessary \citep[see e.g.][]{ChD81}. For that reason we did not consider here the Planck opacity of a 
relatively low magnetized plasma with $B < 10^{11}$\,G.

We also find a relatively simple analytic function which can be used to approximate the numerical computations of the Planck mean opacities:
\be \label{eq:fit}
        \tilde{k}_\parallel = 0.36\,k_0\left(1+A_{\rm cyc}\left(1-\exp{\left[-\frac{k_{\rm B}T}{0.1E'_{\rm cyc}}\right]}\right)^{13.4}\right),
\ee
where $k_0$ is the Planck mean, computed for the non-magnetized plasma at given density and temperature
\be
        k_0 = 0.5058\,\rho T^{-3.5}_{\rm keV},
\ee
and the cyclotron energy $E'$ computed across the field using the relativistic formula
\be 
   E'_{\rm cyc} = m_{\rm e}c^2 \left(\sqrt{1+2\frac{B}{B_{\rm cr}}}-1\right).
\ee
The amplification factor $A_{\rm cyc}$ is also dependent on the magnetic field strength and is
\be
A_{\rm cyc} \approx 4240\,B_{12}^{-1.06}.
\ee
 Here $B_{\rm cr} = 4.414 \times 10^{13}$\,G, see also Appendix \ref{sec:numbers}.

\begin{figure}
\centering
\includegraphics[angle=0,scale=0.99]{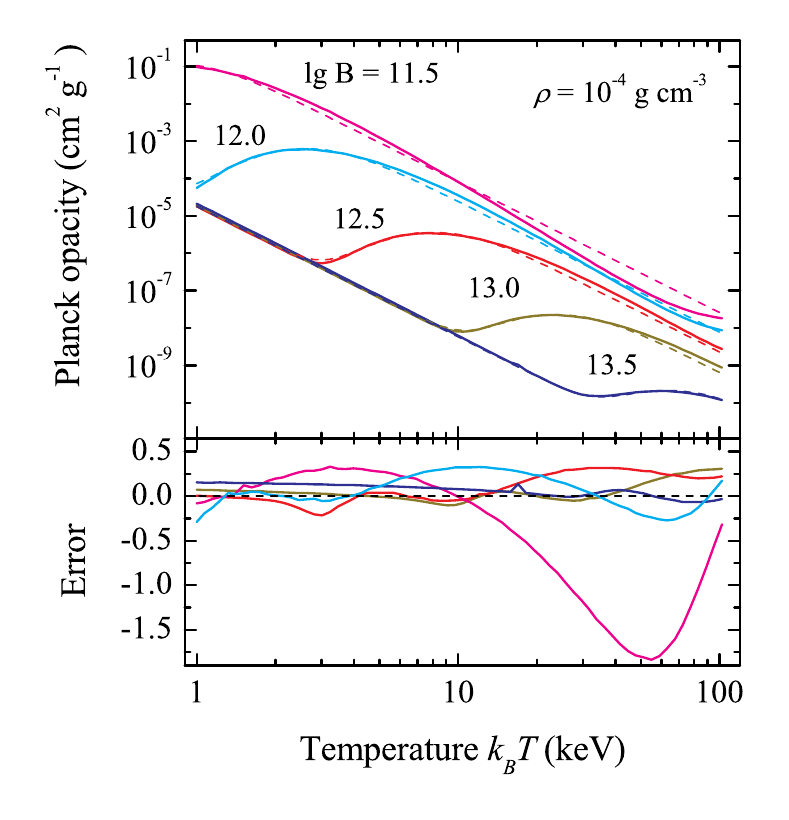}
\caption{\label{fig:p_fit}
{\it Top panel:}  Results of the approximation of the dependencies of the Planck mean opacities on the plasma temperature 
for various magnetic field strengths (solid curves)} with the approximation formula (\ref{eq:fit}, dashed curves). 
 The plasma density is $\rho = 10^{-4}$\,g\,cm$^{-3}$.
{\it Bottom panel:} Relative errors of the fitting.
\end{figure}

Some examples illustrating the parametrization of numerical calculations with this function are shown in Fig.\,\ref{fig:p_fit}. The relative accuracy of the fitting 
$({k}_\parallel-\tilde{k}_\parallel)/{k}_\parallel$ is not high,  about 10-30\%. The error is larger 
for low magnetic fields and can reach 200\% for the lowest magnetic field at high temperatures.
This means that we recommend using the approximation formula (\ref{eq:fit}) with caution for astrophysical problems requiring 
high accuracy.  We note, however, also that the uncertainties due to the simplifications used for numerical calculations, 
especially for the contribution of thermal cyclotron emission $\mathcal P$ could be comparable with the approximation 
formula uncertainties. 

The dependence of the Rosseland mean opacity on plasma density is more complex. Therefore, we computed an extended
grid of Rosseland opacities for both, across and along the field lines
 for 85 plasma temperatures, from $\lg T_{\rm keV} =$\,0 to 2.52 with the step 0.03,  and 14 magnetic field strengths 
 $\lg B =$\,10.5, 11, 11.5, 11.75, 12, 12.25, 12.5, 12.75, 13, 13.25, 13.5, 14, 14.5,  and 15. 
 The third grid parameter is density.  This parameter has 19 values on the grid, from $\lg \rho =$\,$-$6 to 3 with
the step 0.5. We note that the electrons at the lowest plasma temperatures and magnetic field strength are degenerate.
Our approach is not strictly correct for degenerate plasma, and the opacities under such conditions are, therefore, not computed. Instead, the opacity values
computed for the lower plasma density are taken. It is important, that the  magnetic field is strongly
quantizing for the considered plasma parameters at $k_{\rm B}T < E_{\rm cyc}$. 
It means that all the electrons are in the first Landau level 
and the Fermi temperature is significantly less than the Fermi temperature for the non-quantizing magnetic field 
(see Appendix \ref{sec:numbers}). As a result, electrons are not degenerate even at the lowest temperatures and the highest
densities if $\lg B \ge 11.75$.  

\begin{figure}
\centering
\includegraphics[angle=0,scale=0.99]{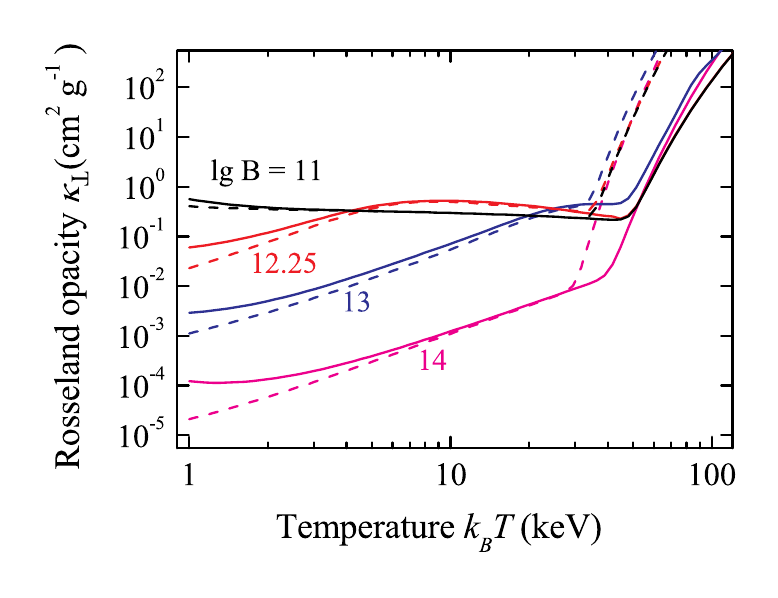}
\caption{\label{fig:rs_op}
Dependencies of the Rosseland mean opacities across the magnetic field  on the plasma temperature for various magnetic field strengths 
and two plasma density parameters,  $\rho = $\,0.1 (dashed curves) and 10\,g\,cm$^{-3}$ (solid curves). The magnetic field strengths
are marked near the curves. 
}
\end{figure}

\begin{figure}
\centering
\includegraphics[angle=0,scale=0.99]{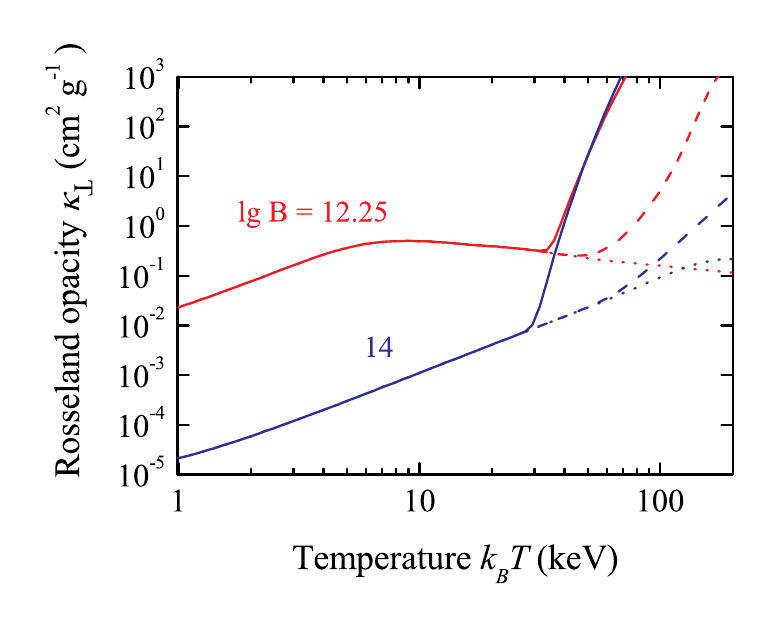}
\caption{\label{fig:no_pair}
  Comparison of the Rosseland $\kappa_\bot$ opacities computed with (solid curves) and without 
(dashed curves) electron-positron pairs taking into account. The opacities computed without considering pairs are also shown (dotted curves)}The results for two magnetic field strengths,
$\lg B =$\,12.25 and $\lg B =$\,14,  and $\rho = 0.1$\,g\,cm$^{-3}$ are shown. 
\end{figure}

\begin{figure}
\centering
\includegraphics[angle=0,scale=0.99]{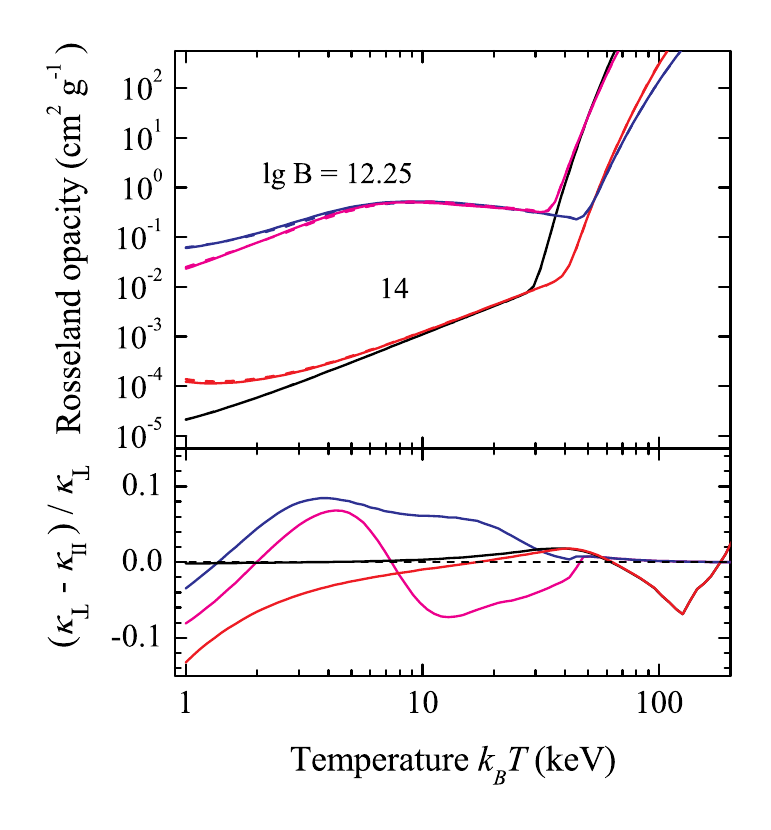}
\caption{\label{fig:rs_par}
Comparison of the Rosseland mean opacities across (solid curves) and along (dashed curves) the magnetic field  
for two field strengths 
and two plasma density parameters,  $\rho = $\,0.1 and 10\,g\,cm$^{-3}$. 
The magnetic field strengths
are marked near the curves. The relative  differences between opacities are shown in the bottom panel.
}
\end{figure}

\begin{figure}
\centering
\includegraphics[angle=0,scale=0.99]{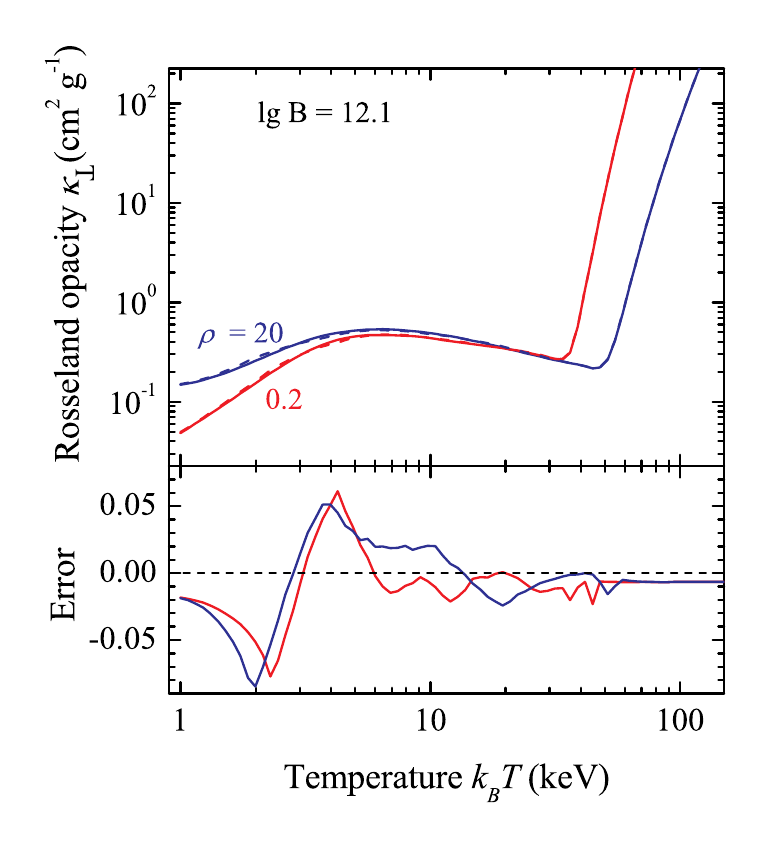}
\caption{\label{fig:rs_int}
{\it Top panel:} Comparison of the interpolated opacities (dashed curves) with the exactly computed ones (solid curves) for
two plasma density parameters,  $\rho = $\,0.2 (red curves) and 20\,g\,cm$^{-3}$ (blue curves). 
The magnetic field strength is fixed 
($\lg B =$\,12.1). {\it Bottom panel:} Dependence of the relative interpolation errors ([exact-interpolated]/exact)
on the plasma temperature.
}
\end{figure}

Examples of the computed Rosseland mean opacities $\kappa_\bot$  from the first set for a few magnetic field strengths 
and two density values,  $\rho = $\,0.1  and 10\,g\,cm$^{-3}$, are shown in Fig.\,\ref{fig:rs_op}. 
A significant increase of the  opacity at $k_{\rm B}T >$40$-$100 keV is connected with the scattering on
 the electron-positron pairs. The opacities at lower temperatures are generally consistent with the case of pure 
 magnetic electron scattering presented by  \citet{Mushtukovetal.15}. The opacities of a dense plasma are higher
than the opacities of a rarified plasma due to increase of the free-free opacity contribution. The opacities for
$B = 10^{11}$\,G  are close to the case of the non-magnetized plasma, and the opacities have a local maximum
at temperatures of about 10 keV for a moderate magnetized plasma ($\lg B =$\,12.25) due to the contribution of 
the cyclotron line and its harmonics.

 A comparison of the Rosseland opacities $\kappa_\bot$ computed for all three considered cases is shown in Fig.\,\ref{fig:no_pair}. 
The opacities from the second case increase at temperatures above 100\,keV mainly due to photon-photon interactions. 
We note, that increase of opacity occurs at the higher temperatures (about 60-70 keV instead of 30 keV in the 
example above), and the opacity is reduced by several orders of magnitude for temperatures above 30 keV.
This computations are not completely self-consistent because the photon-photon interactions lead to 
 pair creation.   However, these results allow us to qualitatively estimate a value of  the Rosseland opacity when the pair number density 
  is far from the equilibrium.  Opacities, computed for the third case, show no significant increase at high temperatures as all the processes connected with pairs are ignored in this set.
 
A comparison of the Rosseland opacity across and along the field is presented in Fig.\,\ref{fig:rs_par}. In fact, the Rosseland 
$\kappa_\bot$ and $\kappa_\parallel$ are close to each other with maximum differences  of about 10-15\%.
The opacities from the first set were used here and further.
 
A  code for the interpolation in the grid was also created.  the code is based on the spline interpolation procedure
MAP1 created by R.L. Kurucz and published in his code ATLAS \citep{Kurucz70}. 
Examples of the relative  interpolation accuracy for $\kappa_\perp$ 
($error = (\kappa_\perp(comp)-\kappa_\perp(interp)/\kappa_\perp(comp)$)
are shown in Fig.\,\ref{fig:rs_int}.  The interpolation accuracy is better than 10\%
for all temperatures. We note that we presented the most complicated case when the contribution of the 
cyclotron opacity is important ($\lg B$ between 11.75 and 12.5). The interpolation accuracy is far better at lower and higher
magnetic fields.  The source files in the arXiv publication contain the interpolation code together with the necessary data files
and a test example.

\section{Conclusions}

We investigated the properties of the Rosseland and Planck mean opacities of a high-temperature plasma in a strong magnetic field. 
Using accurate values of these opacities is important for the construction of accretion structures on the surface of neutron stars in X-ray pulsars, namely accretion columns and accretion heated spots.

We considered a plasma with a simplified chemical composition, the solar hydrogen/helium mix. 
We also assumed that both chemical elements are fully ionized at the considered temperature range $k_{\rm B}T$, 1--330\,keV, and that the plasma density is low enough to be non-degenerate.
Finally, we also assumed that the plasma temperature is larger than  the energy corresponding to the plasma frequency, $k_{\rm B}T \gg E_{\rm p}$.

All specific opacities relevant for radiative transfer in a highly magnetized plasma were considered. 
They included magnetic bremsstrahlung,
magnetic electron scattering, cyclotron line and harmonics, and the one- and two-photon pair creations. 
Magnetic bremsstrahlung and electron scattering were considered in the cold plasma approximations at photon energies $E \le E_{\rm cyc}$, and various simplifying approximations were used for the description of the cyclotron opacity.  
All these three opacity sources were computed for two modes of radiative transfer in a highly magnetized plasma.

We demonstrated that the Planck mean opacity of a highly magnetized plasma has the main properties similar to the Planck mean opacity of the fully ionized non-magnetized plasma. In particular, it is also linearly proportional to the plasma density.
The Planck opacity of the highly magnetized plasma  is by factor of three lower than the corresponding opacity of the 
non-magnetized plasma at temperatures $k_{\rm B}T < 0.1\,E_{\rm cyc}$,
 because only the opacity in the ordinary mode is significant, and the additional reduction arises due to
magnetic Gaunt factor behaviour.  At higher temperatures, the contribution of thermal cyclotron absorption
becomes significant and the Planck opacity increases by a few orders of magnitude reaching a maximum at
 $k_{\rm B}T \approx 0.3\,E_{\rm cyc}$. The opacity amplification is maximum for the lowest magnetic field strength considered
($B \approx 3.3 \times10^{11}$\,G) and decreases almost linearly as the magnetic field strength increases.
 At the same time, the contribution of the  thermal cyclotron emission to plasma cooling is the
largest source of uncertainty for plasma cooling rate, because existing computations of the ratio between the thermal cyclotron emission
and the cyclotron scattering are not robust.
 
 We suggest a relatively simple
approximation formula (\ref{eq:fit}) for the description of the Planck opacity of a high temperature plasma in a strong
magnetic field. We demonstrated that this approximation has an accuracy of about 30\% for $\lg B \ge 12$ 
and can be used for modeling of the radiative cooling of the 
high-magnetized plasma. 

The Rosseland mean opacity does not depend on the plasma density linearly, and we computed an extended grid 
of these opacities both, along and across the field for plasma temperatures from 1 to 330 keV and 14 values of the magnetic field
strength in the range $\lg B$ from 10.5 to 15. The third input parameter of the grid is  the plasma density 
$\rho$ ranging between $10^{-6}$ and 10$^3$. The electrons are degenerate at a few of the lowest temperatures and 
 high densities for low magnetic field strength, and the opacities at these grid points were not computed. 
 The electrons at strongly quantizing magnetic field, $\lg B \ge 11.75$ for the adopted grid are not degenerate even at the lowest
 temperatures and the highest densities.
The main difference of our work in comparison with the calculations previously
 published in the literature is the inclusion of scattering on the electron-positron pairs. Their number densities needed for such 
calculation were computed by us using thermodynamic equilibrium assumption in the non-relativistic approximation.  
Formally, the opacities connected with electron-positron pair creation on the two-photon interaction and the photon interaction
with a strong magnetic field were also included to the Rosseland opacity computations. However, their contribution to the 
total opacity is insignificant in comparison with the scattering on the pairs. 
 Altogether, we computed three sets of Rosseland mean opacities across and along the field. In the first set
we take into account pairs in the thermodynamical equilibrium assumption and opacities due to pair creation in 
the thermodynamic equilibrium radiation field. In the second set we assumed that there are no pairs, but their creation 
in the Planck radiation field is possible. In the last set we ignored the pairs completely.

We  did not include the pair annihilation as 
an additional source of  the radiative cooling of the magnetized plasma, e.g. in the Planck opacity calculations. 
This process can be precomputed assuming that the radiation intensity equals the Planck function, but it is not correct for the
optically thin plasma layers where the Planck opacity has to be  used. 
 
The inclusion of scattering on the pairs
leads to the Rosseland mean opacity dramatically increasing at temperatures $k_{\rm B}T > 50 - 100$\,keV,  
especially for the strong magnetic field cases. This fact refutes earlier results where it was demonstrated that 
the maximum accretion column luminosities increase as the magnetic field strength on the neutron star surface increases
\citep{Mushtukovetal.15, Brice.etal:21}. New computations of the maximum accretion column luminosities with the Rosseland mean opacities computed here taken into account will be published in a separate paper.  Preliminary results
are presented by \citet{2022arXiv220814237S}. There it is shown that the maximum luminosity of the accretion column 
ceases to grow with increasing magnetic field at $\lg B \approx 14.5$, while in the papers cited above the maximum luminosity monotonically grows with increasing magnetic field.

\section*{Acknowledgements}

This work was supported by  Deutsche  Forschungsgemeinschaft  (DFG)  (grant WE 1312/53-1).
AAM thanks UKRI Stephen Hawking fellowship and the Netherlands Organization for Scientific Research Veni Fellowship.

\section*{Data availability}

The Rosseland opacity tables together with the interpolation code will be shared on reasonable request
to the corresponding author. 
They are also available at the arXiv publication  and via \href{https://github.com/alexandermushtukov/RT_mag_opacity}{https://github.com/alexandermushtukov/RT\_mag\_opacity}.


\bibliographystyle{mnras}
\bibliography{allbib}

\appendix

\section{Electron and ion number densities}
\label{sec:numbers}

The electron and ion number densities are determined by 
 two equations, the electric neutrality of the plasma
 \be \label{eq:xe}
  n_{\rm e^-} = n_{\rm e^+}^{\rm B}+ \bar Z\,n_{\rm ion}
\ee
and the relation between the ion number density and the plasma density
\be \label{eq:rho}
n_{\rm ion} = \frac{\rho}{\bar A m_{\rm H}}. 
\ee
 Here  $n_{\rm e^+}^{\rm B}$ is the positron number density at the given magnetic field strength, and this value was taken equal to zero at the Planck opacity computations,
 $\bar Z=A_{\rm H}+2A_{\rm He}$ is the average ion charge, and $\bar A =A_{\rm H}+4A_{\rm He}$ is 
the average ion mass. We take  the relative hydrogen number density $A_{\rm H} = 0.922$ and 
the relative helium  number density  $A_{\rm He} = 0.078$ according to \citet{Asplundetal.09}. 
Therefore, the number densities of hydrogen and helium are computed as follows:
\be
       n_{\rm H} = A_{\rm H} n_{\rm ion}, ~~~~~~n_{\rm He} = A_{\rm He} n_{\rm ion}.
\ee

Note that although there are accurate fully relativistic expressions for the electron and positron number densities in a strong magnetic field 
\citep[see details in][]{Mushtukovetal.19}, we use here the approximation presented by \citet{KYa93}, where the basic assumption is that the electron positron pairs are in thermodynamic equilibrium at the high temperatures.  It means also that we consider a non-degenerate plasma and use non-relativistic approximations. 
This is the situation we look at, so use of the simplified expressions is justified.
 The product of positron and electron number densities in thermodynamic equilibrium 
at zero magnetic field strength can be found as 
\be \label{pos0}
    n_{\rm e^+}^0n_{\rm e^-}^0 = \frac{1}{2\pi^{3}\lambda_{\rm C}^{6}}\, e^{-2/t_{\rm r}}t_{\rm r}^{3},
     ~~~~\lambda_{\rm C}=\frac{\hbar}{m_{\rm e}c},
\ee
\citep{ZN71}. 
It is well known  that the pair number density increases at high magnetic fields when $E_{\rm cyc} \ge m_{\rm e}c^2$,
or $B \ge B_{\rm cr} = m_{\rm e}^2c^3/\hbar e \approx 4.414 \times 10^{13}$\,G \citep[see, e.g.][]{Mushtukovetal.19}.
Here we take into account this kind of amplification using non-relativistic approximation 
\be \label{eq:xp}
          n_{\rm e^+}^{\rm B}n_{\rm e^-} \approx n_{\rm e^+}^0n_{\rm e^-}^0 \Delta b^2(1+0.306\,t_{\rm r})\,\coth^2\left(\frac{\Delta b}{1+2.6\,t_{\rm r}}\right)
\ee 
where
\be
       \Delta b = \frac{E_{\rm cyc}}{2\,k_{\rm B}T}=\frac{\hbar\,eB}{m_{\rm e}c}\cdot\frac{1}{2\,k_{\rm B}T},
\ee
and the relative temperature $t_{\rm r}$ is
\be \label{pe4}
        t_{\rm r} = \frac{k_{\rm B}T}{m_{\rm e}c^2}.
\ee
Finally, using Eqs.(\ref{eq:xe}) and (\ref{eq:xp})
we obtain
\be
    n_{\rm e^+}^{\rm B} = \frac{1}{2}\left(\sqrt{(\bar Z\,n_{\rm ion})^2+4
    n_{\rm e^+}^{\rm B}n_{\rm e^-}}-    \bar Z\,n_{\rm ion}\right). 
\ee
In fact, the plasma density changes due to pair contribution
\be
    \rho' = \rho + m_{\rm e}(n_{\rm e^+}^{\rm B}+n_{\rm e^-}).
\ee
But we consider the plasma density determined using the ion number density only
(Eq.\,\ref{eq:rho}) as a part of the input parameter in our computations, and, therefore, use 
it in opacity computations as well. It means that only the plasma density determined by
Eq.\,(\ref{eq:rho}) has to be used for the Rosseland optical depth computations
\be
    d\tau = \rho\, \kappa(\rho,T,B) dx.
\ee    
We note also, that the opacity calculations presented below can be easily recomputed from
dimension cm$^2$\,g$^{-1}$ used here to dimension cm$^{-1}$ by simple multiplication of opacity values presented here by corresponding grid point density. 

\begin{figure}
\centering
\includegraphics[angle=0,scale=0.99]{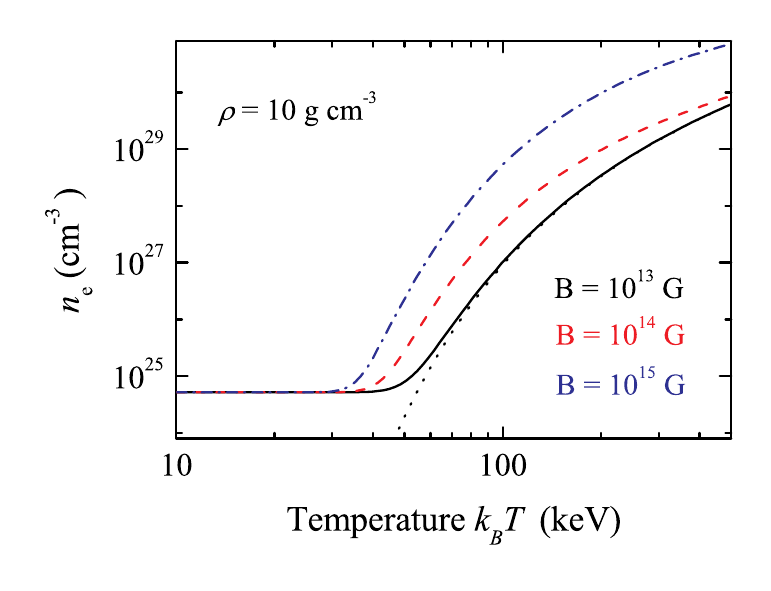}
\caption{\label{fig1}
 The total electron number density ($n_{\rm e^-} + n_{\rm e^+}$) vs. temperature for three magnetic field strengths:
10$^{13}$\,G (black solid curve),   10$^{14}$\,G (red dashed curve), and 10$^{15}$\,G (blue dot-dashed curve).
The plasma density is fixed, $10$\,g\,cm$^{-3}$ . The thermodynamic equilibrium positron number density 
at zero field strength (see Eq.\,\ref{pos0}) is shown with the black dotted curve.
}
\end{figure}

An example of number density calculations using the approximations described below is presented in Fig.\,\ref{fig1}. Here we approximately reproduce 
the bottom panel of the Fig.\,(2) in \citet{Mushtukovetal.19} where more accurate relativistic expressions are used, and as can be seen from the figure, both results are in a good agreement for $k_{\rm B}T < 200$\,keV.

The assumptions adopted in the paper are only appropriate for a non-degenerate plasma. This implies plasma temperatures need to be larger than the Fermi temperature, $T > T_{\rm F}$. The Fermi temperature is determined
by the plasma density and its value is significantly different for a non-quantizing magnetic field, at  $k_{\rm B}T <  E_{\rm cyc}$
and a strongly quantizing magnetic field in the opposite case.

The expression for the Fermi temperature in low field limit is
\be \label{pe6}
        k_{\rm B}T_{\rm e^-}^F = m_{\rm e}c^2(\gamma_{\rm r}-1),
\ee
where the parameters  $\gamma_{\rm r}$ and $x_{\rm r}$ are determined as
\be \label{pe5}
     \gamma_{\rm r} = \sqrt{1+x_{\rm r}^2}, ~~~x_{\rm r} = \lambda_{\rm C}(3\pi^2n_{\rm e^-})^{1/3}.
\ee

In the opposite case, when  $k_{\rm B}T <  E_{\rm cyc}$ we consider the strongly-quantizing approximation,
assuming that only the ground Landau level is occupied \citep{HYaP07}. The Fermi temperature is significantly reduced
at these conditions     
\be \label{pe7}
        k_{\rm B}T_{\rm e^-}^F (B) = m_{\rm e}c^2(\gamma_{\rm B}-1),
\ee
where
\be \label{pe8}
      \gamma_{\rm B} = \sqrt{1+x^2_{\rm B}},~~~~~x_{\rm B} = \frac{2B_{\rm cr}}{3B}x^3_{\rm r}.
\ee

\section{ Specific opacities}
 \label{sopac}

\subsection{Continuum opacity}

The continuum opacity, for both electron scattering and free-free absorption are computed using the method
described by \citet{LH03, vAL06}, see also \citet{SPW09}. In the all cited papers the opacities were computed 
 in cold plasma approximation for 
chemically pure (hydrogen or helium) atmospheres. We modified the method for the H/He mix  and
assume that the cold plasma approximation provides correct opacities at $E < E_{\rm cyc}$. We note that the used
opacities correctly transform to the non-magnetic opacities at $E >> E_{\rm cyc}$. The cyclotron opacity dominates
at $E \sim E_{\rm cyc}$, and the cold plasma approximation is not used  for computations of the cyclotron opacity.

Vacuum polarization is also taken into account according to the description in \citet{vAL06}. It is known that
the plasma polarizability is dominating in the dielectric tensor at photon energies below some boundary energy 
$E_{\rm V}= E_{\rm cyc}\,V^{-1/2}$, where 
\be \label{2}
  V = \frac{3\times 10^{28} {\rm cm^{-3}}}{n_e} \left(\frac{B}{B_{\rm cr}}\right)^4,
\ee
and finally
\be
   E_{\rm V} \approx (1.7 \times 10^4 n_{\rm e})^{1/2}B^{-1}\approx 413 \left(\frac{n_{\rm e}}{10^{25}}\right)^{1/2}
   B_{12}^{-1}\, {\rm keV}.
\ee
In the opposite case, $E > E_{\rm V}$, vacuum polarizability is dominating. There is a so called vacuum resonance 
at the photon energy $E=E_{\rm V}$. At this condition the normal modes are mixed and the opacities in the both modes 
become equal.  

Let us present the formulae for the continuum opacities. The scattering opacity  from the mode $i$ and the direction $\mu$ 
to the mode $j$ and the direction $\mu'$ is
\be
  \kappa^{ij}_{\rm es}(\mu,\mu') =   \frac{3}{4} \frac{\sigma_{\rm T}}{\rho}
\sum_{\alpha=-1}^{+1} B_{\alpha} ~ a_{\alpha}(i,\mu)~a_{\alpha}(j,\mu'),
\ee
and the total scattering opacity from the mode $j$ and the direction $\mu$ is
\be
  \kappa^{j}_{\rm es} (\mu) =   \frac{\sigma_{\rm T}}{\rho} \sum_{\alpha=-1}^{+1}  B_{\alpha}
  ~ a_{\alpha}(j,\mu)~\left(\sum_{i=1}^2 A_{\alpha}(i)~\right),
\ee
where
\be
    B_{\alpha} =n_{\rm e}~ t_{\rm e}(\alpha) +
\left(\frac{m_{\rm e}}{m_{\rm p}}\right)^2 n_{\rm i}\left(A_{\rm H}
t_{\rm H}(\alpha)+A_{\rm He}
t_{\rm He}(\alpha)\right).
\ee
In the described approach the unit mode electric vector in the frame  with the z-axis directed along magnetic field
can be presented as ${\bf E}={\bf e}^j = e_0(iK_j,1,iK_{z,j})$. The vector components are expressed using the ratio 
of the polarization ellipse $iK_j = E_x/E_y$,  its projection on the z-axis   $iK_{z,j} = E_z/E_y$, and the normalization
$e_0 = (1+ K_j^2+K_{z,j}^2)^{-1/2}$. The opacities are computed in the cyclic frame, and the corresponding
squares of the cyclic vector components are 
\bea
   a_{\pm 1}(j,\mu)  &= &\left\vert \frac{1}{\sqrt{2}}(e_x^j+ie_y^j)\right\vert^2 \\ \nonumber
   &= &\frac{1 \pm (K_j \sin
\theta + K_{z,j}\cos\theta)^2}{2(1+K_j^2+K_{z,j}^2)}, 
\eea
\be
   a_{0}(j,\mu)  = \frac{(K_j \sin
\theta-K_{z,j}\cos\theta)^2}{1+K_j^2+K_{z,j}^2},
\ee
and
\be
    A_{\alpha}(j) = \frac{3}{2} \int_0^1 a_{\alpha}(j,\mu)~d\mu.
\ee
They are computed using the ratio of the polarization ellipse axes   
\be
    K_j  =  q\left(1 +(-1)^j\left(1+\frac{r}{q^2}\right)^{1/2}\right), 
\ee
where the polarization parameter $q$
\be
    q = -\frac{(\varepsilon^2-g^2-\varepsilon\eta)\sin^2 \theta +
\varepsilon\eta (1-r)}{2g\eta \cos \theta},
\ee
and projection of the electric field vector on z-axis
\be
    K_{z,j}  =  -\frac{(\varepsilon-\eta)\sin \theta \cos \theta K_j +
g\sin \theta}{\varepsilon \sin^2 \theta + \eta \cos^2 \theta},
 \ee
are computed using the components of the dielectric tensor $\varepsilon$, $g$, and $\eta$.  
We generalize their values, presented by \citet{LH03} after \citet{G70}, for two ions, and
present their shortened expressions without damping terms:
\be
\varepsilon \pm g = 1- \frac{v_{\rm e}+v_{\rm i}}{(1\mp u_{\rm e}^{1/2})(1\pm u_{\rm H}^{1/2})(1\pm u_{\rm He}^{1/2})},
\ee
and
\be
 \eta =1-v_{\rm e}-v_{\rm i}.
\ee  
Here we use the dimensionless values
\be
v_{\rm e}=\frac{E_{\rm p}^2}{E^2};~~~~~~~v_{\rm i}=\frac{E_{\rm p,i}^2}{E^2},
\ee
and
\be
u_{\rm e}=\frac{E_{\rm cyc}^2}{E^2};~~~~~~~u_{\rm H}=\frac{E_{\rm cyc,H}^2}{E^2};
~~~~~~~u_{\rm He}=\frac{E_{\rm cyc,He}^2}{E^2}.
\ee
The positron contribution was taken into account at the plasma photon energy $E_{\rm p}$
computations by replacing  $n_{\rm e} \equiv n_{\rm e^-} + n_{\rm e{^+}}^{\rm B}$ as 
it was declared in Sect.\,\ref{method}.
The energy corresponding to plasma frequency, $E_{\rm p}$, and the electron cyclotron energy $E_{\rm cyc}$ are
also defined in Sect.\,\ref{method}. The energy $E_{\rm p,i}$ corresponds to the ion plasma frequency $\nu_{\rm p,i}^2=
{Z^2e^2 n_i/\pi Am_i}$. The specific contribution of the protons and $\alpha-$particles to the   $\nu_{\rm p,i}$ is
equal, as $Z^2/A \approx 1$ for  both ions. We use the total ion number density
 $n_i = n_{\rm H} + n_{\rm He}$.  The common expression for the ion cyclotron energies is
 \be
 E_{\rm cyc,(H,He)} = \hbar\frac{ZeB}{Am_{\rm p}c} \approx 6.35\cdot 10^{-3}\,B_{12}\frac{Z}{A}\,\rm{keV}. 
\ee

The vacuum polarization changes the dielectric tensor components and they have to be replaced  as
\citep{Potekhinetal.04}
\bea
     \varepsilon~~~~~~~~~&\rightarrow&~~~~~~~~\varepsilon + a' \\ \nonumber
     \eta~~~~~~~~~&\rightarrow&~~~~~~~~\eta + a' + q'.
\eea
The value $r$  is determined as
\be
    r = 1 + \frac{m'}{1+a'} \sin^2 \theta.
\ee
Here $m'$, $a'$ and $q'$ are small corrections expressed as
\bea
     m' & = & -\frac{\alpha_f}{3\pi} \frac{b_Q^2}{3.75+ 2.7 b_Q^{5/4} +
b_Q^2}  \\ \nonumber
     a' & = & -\frac{2\alpha_f}{9\pi} \ln\left(1 +
\frac{b_Q^2}{5}\frac{1+0.25487 b_Q^{3/4}}{1+0.75 b_Q^{5/4}}\right) \\
\nonumber
      q' &=& \frac{7\alpha_f}{45\pi} b_Q^2 \frac{1+1.2 b_Q}{1+1.33 b_Q
+0.56 b_Q^2},
\eea
where $b_Q = B/B_{\rm cr}$, and $\alpha_f= e^2/\hbar c \approx
1/137$.

The magnetic bremsstrahlung opacity in the mode $j$ and the angle $\mu$ separately for collisions of electrons with 
protons and $\alpha -$\,particles is
\be
  \kappa^{j}_{\rm H,He} (\mu) =  \sum_{\alpha=-1}^{+1} \zeta^{j}_{\rm H,He}(\alpha)~ t_{\rm e,(H,He)}
  (\alpha)~a_{\alpha}(j,\mu),
\ee
where
\be
     \zeta^j_{\rm H,He}(\pm 1) = \zeta^0_{\rm H,He} \Lambda_{\bot},~~~~~~~~ \zeta^j_{\rm H,He}(0) = 
     \zeta^0_{\rm H,He}
\Lambda_{\|}.
\ee
Here $\Lambda_{\bot}$ and $\Lambda_{\|}$ are the magnetic Gaunt factors \citep{PP76, SPavW10,Potekhin10}, 
and $\zeta_{0}$ is the nonmagnetic bremsstrahlung opacity without Gaunt factor 
\be
\zeta^0_{\rm H,He} = \frac{2^{5/2}\pi^{3/2}\hbar^2 e^6}{c\, m_{\rm
e}^{3/2}(kT)^{1/2}}  \frac{ Z^2A_{\rm H,He}n_{\rm
i}n_{\rm e}}{E^3\rho}~\left(1-e^{-\frac{E}{kT}}\right).
\ee

We also use the following definitions
\be
  t_{\rm e}(\pm 1) =  \frac{1}{(1 \pm u_{e}^{1/2})^2 +
(\gamma_{\rm e,H}^{\bot} +\gamma_{\rm e,He}^{\bot} + \gamma_{re})^2}, 
\ee
\be
  t_{\rm H,He}(\pm 1)  =  \frac{1}{(1 \mp u_{\rm H,He}^{1/2})^2 +
(\gamma_{\rm e,(H,He)}^{\bot} + \gamma_{ri})^2},  
\ee
\be
  t_{\rm e,(H,He)}(\pm 1)  =  \frac{1}{(1 \pm u_{e}^{1/2})^2(1 \mp
u_{\rm H.He}^{1/2})^2 + \Gamma_{\rm e,(H,He)}},  
\ee
and
\be
 t_{\rm e}(0) =  t_{\rm H,He}(0)  = t_{\rm e(H,He)}(0)  =  1.
\ee
Here
\be
\Gamma_{\rm e,(H,He)} = \left(\gamma_{\rm e,(H,He)}^{\bot} + (1 \pm u_{e}^{1/2})\gamma_{ri}+
 (1 \mp u_{\rm H,He}^{1/2})\gamma_{re}\right)^2.
\ee

In the previous equations we used the effective radiative dimensionless dumping rates 
\be
  \gamma_{re} = \frac{2e^2}{3m_{\rm e}c^3\hbar}\,E,~~~~~~\gamma_{ri} = \frac{2Z^2e^2}{3Am_{\rm p}c^3\hbar}\,E.
\ee
We note, that the rates for protons and $\alpha -$\,particles are equal to each other,
and we do not distinguish them. The relative effective dimensionless electron-ion collision rate across the magnetic field is
\be
\gamma_{\rm e,(H,He)}^{\bot} = \frac{\sqrt{2\pi}\hbar\, Z^2e^4}{(m_{\rm e}kT)^{1/2}}\frac{A_{\rm H,He}n_{\rm i}}{E^2}
~\left(1-e^{-\frac{E}{kT}}\right)\, \Lambda_{\bot}.
\ee

\begin{figure}
\centering
\includegraphics[angle=0,scale=0.99]{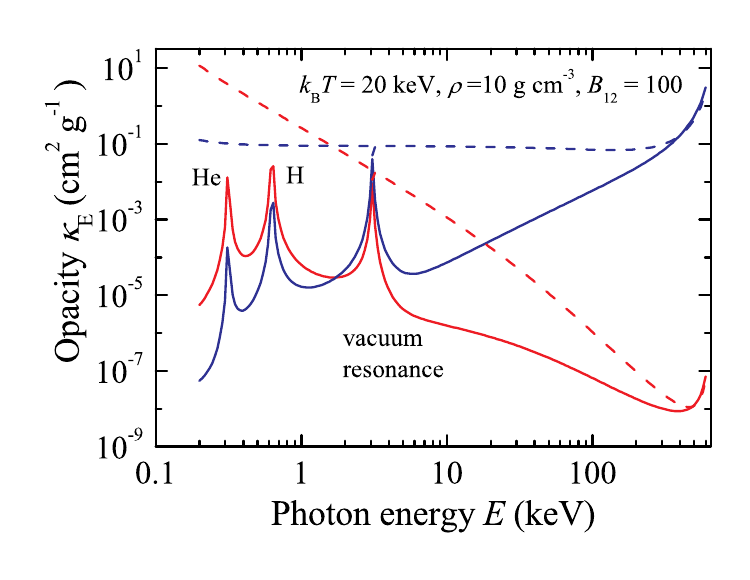}
\caption{\label{fig2}
Dependence of the opacities due to free-free absorption (red curves) and electron scattering opacities (blue curves) 
on the photon energy for a plasma with the parameters presented in the plot. The magnetic field strength is 10$^{14}$\,G. 
Opacities in X-mode are shown by the solid curves, and the opacities in O-mode are
shown by the dashed curves. Positions of the ion resonances for protons and $\alpha-$particles as well as the position 
of the vacuum resonance are also shown.
}
\end{figure}

Examples of continuum opacities for some plasma parameters are shown in Fig.\,\ref{fig2}.
We ignore here bremsstrahlung opacities due to $e^-e^-$,  $e^+e^+$, and $e^-e^+$ which could be
potentially important at high magnetic fields and temperatures. We believe that the Rosseland mean opacity is
dominated by photon scattering by electrons and positrons.

The electron scattering opacity can be smaller than Thomson scattering due to the Klein-Nishina reduction.
It could be significant at high plasma temperatures. Therefore, we multiply the electron scattering opacity
by the known ratio $\sigma_{\rm KN}/\sigma_{\rm T}$ at every photon energy. 
Here $\sigma_{\rm T}=6.65\times10^{-25}$\,cm$^2$ is the Thomson scattering cross-section. We perform 
this kind of reduction for the cyclotron opacity as well.
The importance of this reduction for the Rosseland opacity of a weakly magnetized plasma 
is even more significant if Compton scattering
is taken into account, see e.g. \cite{P83,SWP12,P17}, but we ignore it here.

\subsection{Cyclotron opacity}

The opacities in the cyclotron line and the harmonics $\kappa_j$ (cm$^2$ g$^{-1})$ 
for polarization modes 
$j$=1 (X-mode) and $j=2$ (O-mode) are computed using various approximate expressions.

For the normal mode opacities for the plasma domination case we use the approximate formulae 
derived by  \citet{PSY80}, \citep[see also][]{SPW12}.
In particular, the X-mode opacity near the fundamental resonance is
\be \label{u4}
  \kappa_{11} \approx  \frac{ \sqrt{\pi}}{\hbar c\rho} 
\frac{E_p^2}{E}
\frac{1+\mu^2}{2\beta_{\rm T} |\mu|}\exp(-x^2_1).
\ee
The expressions for the O-mode opacity near the fundamental resonance as well as the opacities at the harmonics
can be found in \citet{SPW12}.

In comparison with the original version we have introduced some relativistic corrections by hands.  So we use another
definition for $x_1$  and corresponding values for higher harmonics
\be
      x_s = \frac{E-E_{\rm s}}{\Delta E_{\rm D,s}}
\ee
by including the correct relativistic expression for the resonance and harmonic energies $E_{\rm s}$ instead of 
using the approximate linear shift relative to $sE_{\rm cyc}$
\be \label{Ecyc}
       E_{\rm s} =\frac{m_{\rm e}c^2}{1-\mu^2}\left(\sqrt{1+2sB(1-\mu^2)/B_{\rm cr}}-1\right).
\ee
Here $s$ is the harmonic number, and $s=1$ for the fundamental resonance. The expression (\ref{Ecyc}) is correct for
$\mu \neq 1$ only. There is only fundamental resonance at the energy $E_{\rm cyc}$ at $\mu =1$.

The Doppler width of the harmonic $\Delta E_{\rm D,s}$ is also corrected,
\be
        \Delta E_{\rm D,s} = \min(E, E_{\rm s})\,\beta_{\rm T} \mu 
        \left(\frac{\sqrt{1-\beta_{\rm T}^2}}{1-\mu\beta_{\rm T}}\right).
\ee
Here $\beta_{\rm T}$ is the ratio of the most probable thermal electron velocity to the speed of light. It is also corrected
for relativistic effects,
\be
        \beta_{\rm T} =\frac{\vv_{\rm T}}{c} = \sqrt{2t_{\rm r}}\left(\sqrt{1+t_{\rm r}^2}-t_{\rm r}\right)^{1/2},
\ee
where $t_{\rm r}$ is defined by Eq.\,(\ref{pe4}).This expression provides the requirement  $\beta_{\rm T}<1$ 
for any temperature, and  can be derived from the condition 
$\partial f(p)/\partial p = 0 $, where $f(p) \sim p^2 \exp(-\sqrt{1+p^2}/t_{\rm r})$ is the probability distribution of electron
momenta in accordance with the relativistic Maxwell-J\"uttner 
distribution, 
and $p=|\vv|/c\sqrt{1-\vv^2/c^2}$ is  the module of the  dimensionless relativistic momentum. 

In expression (\ref{u4}) as well as in all the following expressions for the cyclotron opacities, the Gaussian broadening function
is used instead of integration over the electron momenta. It is acceptable for $t_{\rm r} \ll 1$, but for the high temperature 
cases the resonance condition is not fulfilled for the high energy electrons with the momenta directed to the observer.
Therefore, we cut the cyclotron opacity at energies higher than the cutoff energy 
\be
       E_{\rm cut,s} =\frac{m_{\rm e}c^2}{\sqrt{1-\mu^2}}\left(\sqrt{1+2sB/B_{\rm cr}}-1\right)
\ee
which is determined separately for the fundamental resonance and every harmonic, see, e.g. \citet{Schwarmetal.17}.

For the case of the vacuum domination we also use the approximations for the opacities near the fundamental resonance
from \citet{PSY80}. The opacity at the fundamental resonance in X-mode is
\be \label{u5}
  \kappa_{11}   \approx  \frac{ \sqrt{\pi}}{\hbar c\rho} 
\frac{E_p^2}{E}
\frac{1}{2\beta_{\rm T} |\mu|}\exp(-x^2_1).
\ee
This expression is correct, if the value  $V\beta_{\rm T}\mu(1-\mu^2) > 1$ \citep[see detail in][]{PSY80}.
In the opposite case we use Eq.\,(\ref{u4}). The same condition we use for separation of the expressions describing 
the opacities near the fundamental resonance in O-mode.

The opacities near the harmonics in X-mode for the case of the vacuum dominance are taken from \citet{MZh81} 
\be \label{u7}
  \kappa_{s1} \approx  \frac{\sqrt{\pi}}{\hbar c\rho} 
 \frac{E_p^2}{E}\frac{1}{\beta_{\rm T} |\mu|}\frac{\kappa_V}{(s-1)!}\, \exp(-x^2_s),
\ee
where
\be
      \kappa_V = 
      \frac{\sinh(E/2k_{\rm B}T)}{(\exp(E_{1}/k_{\rm B}T))-1)^s}
      \left(\frac{\varepsilon E(1-\mu^2)}{2E_1}\right)^{s-1}.
\ee
Here  $\varepsilon=E/m_{\rm e}c^2$ is the relative photon energy.
These expressions are correct for the case $k_{\rm B}T \sim E_{\rm cyc}$. The cyclotron opacity is important
for the Rosseland opacity computation in this case only.

All the opacities near the fundamental resonance and the harmonics in O-mode for the case of the vacuum domination 
are computed using the expressions for the opacities in X-mode 
\be \label{u6}
  \kappa_{s2} \approx  \kappa_{s1} \left(\mu^2  + \frac{E_1}{4k_{\rm B}T}\beta_{\rm T}^2\right),
\ee
but the separation condition for the harmonics is the same as for continuum opacities, i.e. we use Eq.\,(\ref{u6}) if $E>E_{\rm V}$. 
In fact the transition of the cyclotron opacity across the vacuum resonance is much more complicated 
\citep[see, e.g. ][]{PSY80} and we did not consider it here in detail. 

\begin{figure}
\centering
\includegraphics[angle=0,scale=0.99]{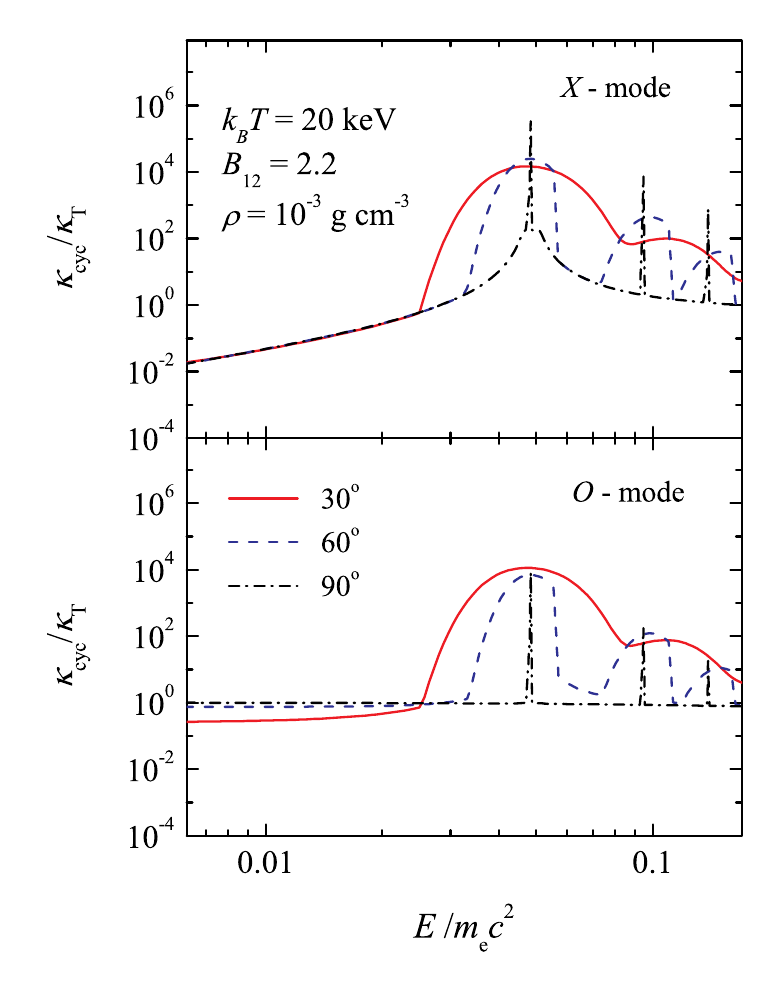}
\caption{\label{fig3}
Dependence of the cyclotron plus continuum electron scattering opacities on the relative photon energy for 
a plasma with temperature $k_{\rm B}T=20$\,keV, $\rho =10^{-3}$\,g\,cm$^{-3}$, and  $B=2.2.\times 10^{12}$\,G. 
All the opacities are computed for the vacuum domination case.
}
\end{figure}

Examples of cyclotron opacities are presented in Fig.\,\ref{fig3}. It reproduces Figs.\,5a and 6a published
by \citet{2016PhRvD..93j5003M}. It is clear that our simplified approach reproduces well the resonance and harmonic energies,
although the line widths are slightly underestimated (especially for $\mu \rightarrow 0$) because we ignore the natural Landau 
level width, and the opacities in the harmonics are overestimated. The cyclotron opacities are presented together with the 
continuum electron scattering opacities at low energies.

All the expressions presented above
are not sufficiently correct at high temperatures and  using of them
leads to cyclotron opacity overestimation. This is the reason why we do not use them for
the cyclotron harmonic opacities at temperatures above 45 keV and the large angles provide too broad  harmonics
for the plasma domination case.
At these conditions we use the opacities derived from the fitting formulae for the high-temperature cyclotron-synchrotron
emissivities obtained by \citet{Pandyaetal.16}:
\be \label{syn}
       \kappa^j_{\rm syn} = \frac{k_{\rm syn}}{\rho}(X \pm Y),
\ee
where the sign $"+"$ corresponds to X-mode, and the sign $"-"$ corresponds to O-mode.
Here
\be
   k_{\rm syn} = \frac{\sqrt{2}\pi}{27}e^2 n_{\rm e}\sqrt{1-\mu^2}\, E_{\rm cyc}\exp(-\epsilon^{1/3})B_E^{-1},
\ee
\be
X = (\epsilon^{1/2} + 2^{11/12}\epsilon^{1/6})^2, 
\ee 
and
\be
Y = (\epsilon^{1/2} + a_{\rm t} 2^{11/12}\epsilon^{1/6})^2,
\ee 
where
\be
   a_{\rm t} = \frac{7 t_{\rm r}^{24/25}+35}{10 t_{\rm r}^{24/25}+75}.
\ee
The dimensionless relative energy $\epsilon = E/E_{\rm c}$ is the photon energy normalized by the energy
\be
  E_{\rm c} = \frac{2}{9} E_{\rm cyc} t_{\rm r}^2 \sqrt{1-\mu^2},
\ee
and $B_E$ is the Planck function.
The opacity determined by Eq.\,(\ref{syn}) is used as the cyclotron opacity of the harmonic if $k_{\rm B}T > 45$\, keV and 
the plasma domination case.
The opacities near the fundamental resonance and all the opacities for the case of the vacuum dominance are computed 
using Eqs.(\ref{u4}), (\ref{u5}), (\ref{u7}), and (\ref{u6}) for any plasma temperature.

\subsection{Two-photon pair production}

Interaction of two photons with energies $E$ and $E'$ can create an electron-positron pair, 
if their  common relative energy  in the centre-of-momentum frame 
 exceeds the pair mass, $\varepsilon \varepsilon'(1-\mu') > 2$, where $\mu'$ is the cosine of the angle between  the directions 
 of the photon propagations, and  $\varepsilon = E/m_{\rm e}c^2$ is the relative photon energy. 
 The corresponding opacity is described in detail by \citet{B99}. We simplified his expressions as we
 considered pair creation by the almost isotropic blackbody radiation. As a result the opacity is 
\be \label{posop1}
   \kappa_{\gamma\gamma} (E)  =\frac{2\pi}{c\rho} \int\limits_{0}^{\infty}\,\frac{B_{E'}\,dE'}{E'}
    \int\limits_{-1}^{+1}\,
     (1-\mu' )\, \sigma_{\gamma\gamma}(\varepsilon_{\rm c})\,d\mu' .
\ee
Here  $\sigma_{\gamma \gamma}(\varepsilon_{\rm c})$ is the cross-section of the two-photon interaction 
\begin{eqnarray}
\sigma_{\gamma\gamma}(\varepsilon_{\rm c}) =  \frac{3 \sigma_{\rm T}}{8 \varepsilon_{\rm c}^2} 
\left(2+\frac{2}{\varepsilon_{\rm c}^2}+ \frac{1}{\varepsilon_{\rm c}^4}\right)\ln\left(\varepsilon_{\rm c}+
\sqrt{\varepsilon_{\rm c}^2 - 1}\right)  \\ \nonumber
-\frac{3 \sigma_{\rm T}}{8 \varepsilon_{\rm c}^2}\left(1+\frac{1}{\varepsilon_{\rm c}^2}\right)
\sqrt{1-\frac{1}{\varepsilon_{\rm c}^2}}, 
\end{eqnarray}
depending on the relative photon energy  in the centre-of-momentum $z_{\rm c}$
\be
       \varepsilon_{\rm c} = (\varepsilon'/\varepsilon_{\rm thr})^{1/2},
\ee
where
\be
     \varepsilon_{\rm thr} = \frac{2}{\varepsilon(1-\mu' )}.
\ee
The cross-section $\sigma_{\gamma \gamma}(\varepsilon_{\rm c})$ equals zero if $\varepsilon' <\varepsilon_{\rm thr}$.

\begin{figure}
\centering
\includegraphics[angle=0,scale=0.99]{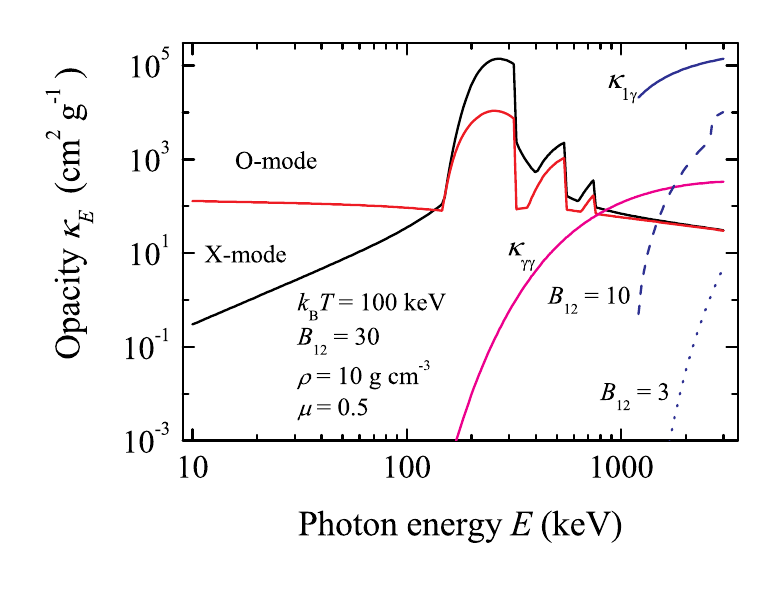}
\caption{\label{fig4}
Dependence of the cyclotron plus continuum electron scattering opacities on the relative photon energy for 
a plasma with temperature $k_{\rm B}T=20$\,keV, $\rho =10^{-3}$\,g\,cm$^{-3}$, and  $B=2.2.\times 10^{12}$\,G. 
All the opacities are computed for the vacuum domination case.
}
\end{figure}

\subsection{One-photon pair production in a strong magnetic field}

In a strong magnetic field a single photon with energy $\varepsilon_\perp \equiv \varepsilon \sqrt{1-\mu^2} > 2$ 
can transform to an electron-positron pair.
We consider here this process as an additional opacity, and use two approximate expressions for the attenuation  rate.
The first one was presented by \citet{DH83} and we use it for the strong enough magnetic field $B > 0.5 B_{\rm cr}$.
 If the photon energy is larger than the $\varepsilon_\perp > 2$, then 
\be \label{1gam}
       \kappa_{1\gamma} (\mu, E)\approx 0.23 \frac{\alpha_f}{\lambda_{\rm C}} \frac{B\sqrt{1-\mu^2}}{B_{\rm cr}}
       \exp\left[-\frac{4f(E,B)}{3\chi}\right] \,\frac{1}{\rho}.
\ee
In the opposite case, $\varepsilon_\perp< 2$, the considered opacity equals zero, $\kappa_{1\gamma} = 0$.
Here the variable $\chi$ is determined as
\be       
       \chi = \frac{\varepsilon_\perp }{2} \frac{B}{B_{\rm cr}},
\ee
the fitting function is
\be
     f(E,B) \approx 1+ 0.42\left(\frac{\varepsilon_\perp }{2}\right)^{-2.7}
      \left(\frac{B}{B_{\rm cr}}\right)^{-0.0038},
\ee
and $\alpha_f \approx 1/137$ is the fine-structure constant. The presented one-photon annihilation opacity is averaged
over the photon polarizations and over the photon energy, smearing the numerous saw-edges existing in the accurate 
attenuation rate. 

At relatively low magnetic fields Eq.\,(\ref{1gam}) significantly overestimates the correct pair creation rate. For this case another
approximation, suggested by \citet{Baring91}, is used:
\be \label{1gam1}
       \kappa_{1\gamma} (\mu, E)\approx  \frac{\alpha_f}{\lambda_{\rm C}} \frac{B\sqrt{1-\mu^2}}
       {\varepsilon^2_\perp B_{\rm cr}}\Lambda(\varepsilon_\perp)
       \exp\left[-\frac{\phi B_{\rm cr}}{4B}\right] \,\frac{1}{\rho}.
\ee
Here
\be
   \Lambda(\varepsilon_\perp) = \frac{3\varepsilon^2_\perp-4}{(\varepsilon_\perp+2)^2}
   \sqrt{\frac{\varepsilon^2_\perp-4}{\zeta\phi}},
\ee
\be
     \phi = 4\varepsilon_\perp-\zeta(\varepsilon^2_\perp-4),
\ee
and
\be
        \zeta = \log\left(\frac{\varepsilon_\perp+2}{\varepsilon_\perp-2}\right).
\ee
Examples of the $\kappa_{1\gamma} (\mu, E)$ dependence on photon energy for three different magnetic field strengths
are shown in Fig.\,\ref{fig4}.  

\section{Data files and the interpolation code}
\label{code}

The arXiv ``Source Files" contain the files with the grid Rosseland opacities across and along  magnetic field lines.
The opacity database is also available via \href{https://github.com/alexandermushtukov/RT_mag_opacity}{https://github.com/alexandermushtukov/RT\_mag\_opacity}.
The opacities represented in different files are computed under different assumptions about contribution of electron-positron pair. 
 In particular, the files contained in archives {\sf ross1.zip} and {\sf ross2.zip} correspond to the opacities across and along the field  respectively  with the pairs in thermodynamic equilibrium and opacities due to pair creation taken into account. 
The  files contained in archives {\sf ross3.zip} and {\sf ross4.zip} correspond to the opacities across and along the field  respectively  with no pairs, but the opacities due to pair creation taken into account. 
 The files contained in archives {\sf ross5.zip} and {\sf ross6.zip} correspond to the opacities across and along the field respectively computed without pairs participation at all.

Every archive file contains information about the Rosseland opacities calculated
 for 14 values of magnetic fields $\lg B$\,= 10.5, 11.0, 11.5, 11.75, 12.0, 12.25,
12.5, 12.75, 13.0, 13.25, 13.5, 14.0, 14.5, 15.0, where $B$ is measured in Gauss, for 19 values of mass density, from $\lg \rho$\,= -6 to 3 with
the step 0.5, and  for 85 values of temperature, from $\lg T$\,=0 to 2.52 with the step 0.03.

Every archive file is organised as a set of 14 two-dimensional tables 
of the opacities collected in separate data files named like
{\sf ross1\_b1175.dat}. 
Every two-dimensional table is computed for the magnetic field strength marked in the file name.
In the example above, the magnetic field strength corresponds to $\lg B$\,= 11.75. 
Each table  consists  of 85 rows corresponding to the grid temperatures and 21 columns. 
In the first column, the temperature in units of keV is presented, in the second  column, the values $\lg T$ are presented, and in the rest columns, the values of the Rosseland mean opacities for 
19 values of the plasma density are presented, starting with $\lg \rho$\,=$-$6 (third column).

We also provide  an interpolation code  finding the Rosseland opacity {for given}  values of $B$, $T$, and $\rho$ inside of the grid, and a test code. 
Both codes  are in  file {\sf ross\_intSn.f}. 
Before use, all the data files have to be unpacked into 
the same directory with the code (or possibly in a separate directory, but in this case, the accurate ways to them have to be determined by  {\sf open()} operators).

The interpolation subroutine {\sf abross1(istart,rho,ttt,bb,abrss)}
has three  input parameters 
{\sf bb}$\equiv B$(G), 
{\sf ttt}$\equiv k_{\rm B} T$(keV),
{\sf rho}$\equiv \rho$(g\,cm$^{-3}$), 
 one output parameter {\sf abross}$\equiv \kappa_{\rm R}$, 
and one key {\sf istart}.
Before interpolation, a choice of the type of the necessary Rosseland opacities has to be made, which is coded as {\sf rossN}.
For this aim, the subroutine {\sf abross1} has to be called with the key {\sf istart}=$N$ and arbitrary values of other input parameters.
Then the necessary interpolation can be performed by calling the subroutine with {\sf istart}=0 and the actual values 
of the input parameters. 
It is possible to make as many opacity interpolations as it is necessary after the establishment of some Rosseland opacity type. 
A new call of the subroutine with another {\sf istart}=$N'$ has to be performed if it is necessary to find another Rosseland opacity type, coded as {\sf rossN'}.

\end{document}